\documentclass[3p,times,procedia,superscriptaddress]{elsarticle}
\usepackage{nupha_ecrc,color}
\usepackage{graphicx}% Include figure files

\volume{00}

\firstpage{1}

\journalname{Nuclear Physics A}

\runauth{}

\jid{nupha}

\jnltitlelogo{Nuclear Physics A}

\usepackage{amssymb}
\usepackage[figuresright]{rotating}

\begin{document}

\begin{frontmatter}

%% Title, authors and addresses

%% use the tnoteref command within \title for footnotes;
%% use the tnotetext command for the associated footnote;
%% use the fnref command within \author or \address for footnotes;
%% use the fntext command for the associated footnote;
%% use the corref command within \author for corresponding author footnotes;
%% use the cortext command for the associated footnote;
%% use the ead command for the email address,
%% and the form \ead[url] for the home page:
%%
%% \title{Title\tnoteref{label1}}
%% \tnotetext[label1]{}
%% \author{Name\corref{cor1}\fnref{label2}}
%% \ead{email address}
%% \ead[url]{home page}
%% \fntext[label2]{}
%% \cortext[cor1]{}
%% \address{Address\fnref{label3}}
%% \fntext[label3]{}

%% Instructions from Editor: Please use the following \dochead only in the preprint version (e-print arXiv etc.); 
%% use empty \dochead{} when submitting to Nuclear Physics A!
\dochead{XXVIth International Conference on Ultrarelativistic Nucleus-Nucleus Collisions\\ (Quark Matter 2017)}
%\dochead{}
%% Use \dochead if there is an article header, e.g. \dochead{Short communication}
%% \dochead can also be used to include a conference title, if directed by the editors
%% e.g. \dochead{17th International Conference on Dynamical Processes in Excited States of Solids}

\title{Phase transitions in dense matter}

%% use optional labels to link authors explicitly to addresses:
%% \author[label1,label2]{<author name>}
%% \address[label1]{<address>}
%% \address[label2]{<address>}

\author[label1]{Veronica Dexheimer}
\address[label1]{Department of Physics, Kent State University, Kent OH 44242 USA}

\author[label2]{Matthias Hempel}
\address[label2]{Department of Physics, University of Basel, Basel, Switzerland}

\author[label3,label4]{Igor Iosilevskiy}
\address[label3]{Joint Institute for High Temperature of RAS, Moscow, Russia}
\address[label4]{Moscow Institute of Physics \& Technology (State University), Moscow, Russia}

\author[label5]{Stefan Schramm}
\address[label5]{FIAS, Johann Wolfgang Goethe University, Frankfurt, Germany}

\begin{abstract}
As the density of matter increases, atomic nuclei disintegrate into nucleons and, eventually, the nucleons themselves disintegrate into quarks. The phase transitions (PT's) between these phases can vary from steep first order to smooth crossovers, depending on certain conditions. First-order PT's with more than one globally conserved charge, so-called non-congruent PT's, have characteristic differences compared to congruent PT's. In this conference proceeding we discuss the non-congruence of the quark deconfinement PT at high densities and/or temperatures relevant for heavy-ion collisions, neutron stars, proto-neutron stars, supernova explosions, and compact-star mergers.
\end{abstract}

\begin{keyword}
%% keywords here, in the form: keyword \sep keyword
QCD phase diagram \sep phase transition \sep quark deconfinement \sep isospin asymmetric matter
%% MSC codes here, in the form: \MSC code \sep code
%% or \MSC[2008] code \sep code (2000 is the default)
\end{keyword}

\end{frontmatter}

%%
%% Start line numbering here if you want
%%
% \linenumbers

%% main text
%\section{}
%\label{}

%% The Appendices part is started with the command \appendix;
%% appendix sections are then done as normal sections
%% \appendix

%% \section{}
%% \label{}
%\section{{\color{red} Remove} Introduction}
{\color{white}.}\\* 

The formalism necessary to describe hadronic and quark matter at high densities and/or temperatures is still an open question and also one of the main goals of current nuclear physics research. Since current lattice QCD calculations cannot reach the regime of high densities due to the highly oscillatory behavior in the functional integral, it is not possible to describe the equation of state of dense, strongly interacting matter on a fundamental level. However, assuming that only baryon degrees of freedom are relevant for intermediate energy scales, the nuclear interaction can be reasonably approximated by effective relativistic mean field hadronic models. In such approaches, the baryon-baryon interaction is described by the exchange of scalar and vector mesons, which simulate the attractive and repulsive features of the nuclear interaction. 

In addition, for higher energy scales, features such as chiral symmetry restoration and deconfinement to quark matter must be reproduced, together with being in agreement with perturbative QCD. In order to investigate general features of the QCD phase diagram, we make use of the Chiral Mean Field (CMF) model \cite{Dexheimer:2009hi}. It is an extended non-linear realization of the SU(3) sigma model, which uses pseudo-scalar mesons as parameters of chiral transformation. It includes the baryon octet, leptons, and quarks as degrees of freedom and it was fitted to reproduce nuclear, lattice QCD, heavy-ion collision, and astrophysical constraints. The degrees of freedom change from hadrons to quarks through a contribution of the field $\Phi$ (in analogy to the Polyakov loop) in their effective masses. A potential for $\Phi$ present at all chemical potentials and temperatures generates the first-order phase transition (PT) coexistence lines seen in Fig.~1. 

Figure~1 highlights the fact that there is no fundamental reason to make use of different descriptions or models to describe astrophysical objects and the energetic matter produced in laboratories. With the realization that astrophysical events such as supernova explosions and neutron-star mergers can reach temperatures of $T=30$ MeV \cite{Burrows:1986me,Pons:1998mm,Pons:2001ar} and $T=80$ MeV \cite{Bauswein:2012ya}, respectively, and heavy-ion experiments like HADES, FAIR, NICA, and the beam energy scan at RHIC can reach large densities, there is no fundamental separation between those two kinds of systems. In Fig.~1, one can see a first-order PT coexistence line corresponding to the nuclear liquid-gas PT for isospin-symmetric matter at low chemical potentials and temperatures (reproduced within the CMF model). For large chemical potentials and temperatures, one finds a deconfinement line for isospin-symmetric matter (with no net strangeness) and one for neutron-star matter (charge neutral and chemically equilibrated), again calculated within the CMF model. All coexistence lines end up in critical points, beyond which the transitions become smooth crossovers. It is important to point out that we obtain the deconfinement critical points under the assumption that, for finite temperature, there are quarks in the hadronic phase and hadrons in the quark phase and each phase is defined by the value of the order parameter for deconfinement $\Phi$.

\begin{figure}[t!]
\begin{center}
\includegraphics[width=0.74\textwidth]{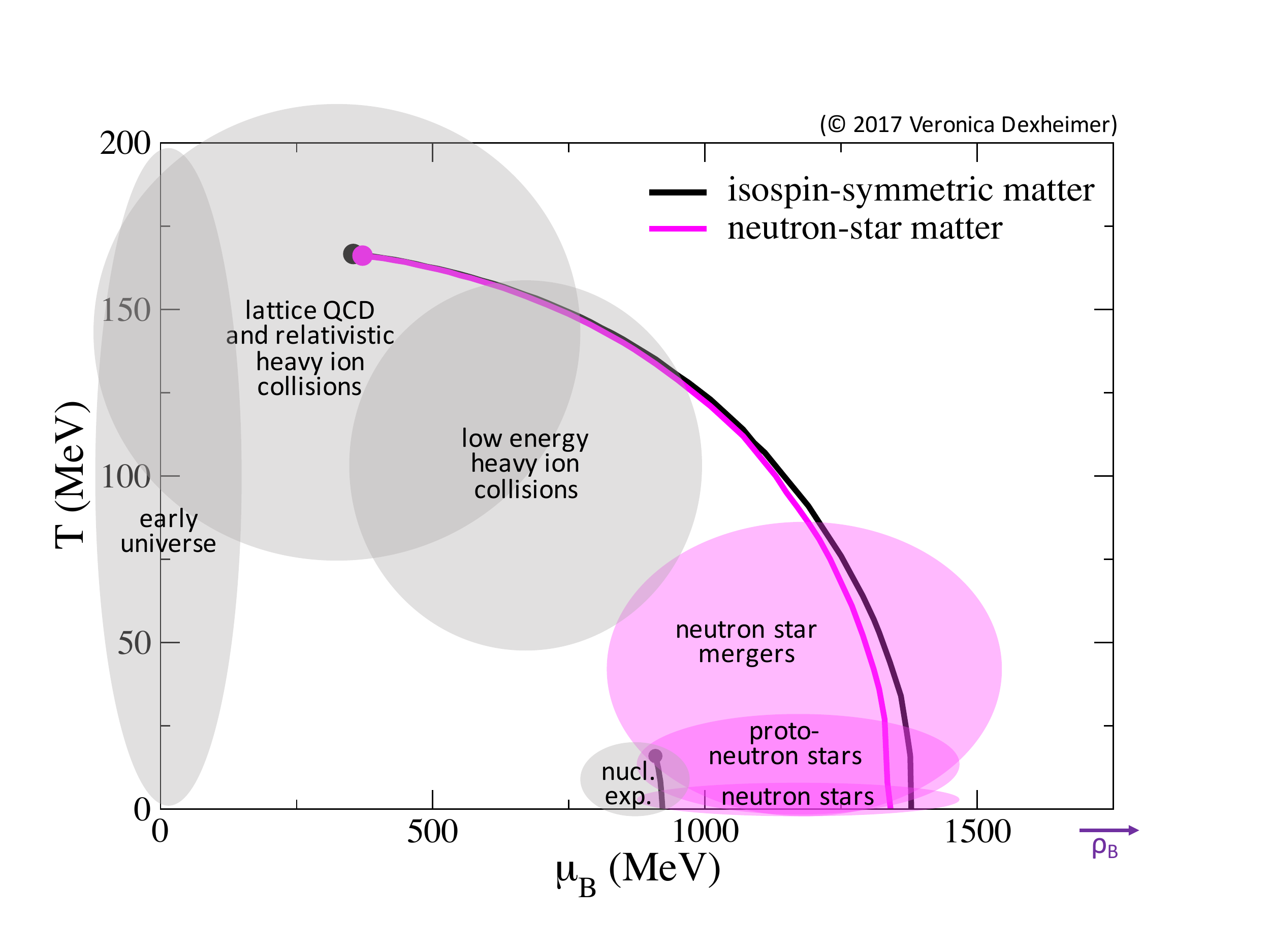}
\caption{(Color online) QCD Phase diagram for isospin-symmetric matter (with zero net strangeness) and neutron-star matter (charge neutral and in chemical equilibrium) calculated using the CMF model.}
\end{center}
\end{figure}

PT's in systems with 2 macroscopic phases that possess more than one globally conserved charge are of non-congruent type (see Refs.~\cite{Iosilevskiy:2010qr,Hempel:2013tfa,Hempel:2015eoj} and references therein for details). In the isospin-symmetric case, this happens due to baryon number and isospin (or electric charge). Usually, in non-congruent PT's, the local concentration of the charge associated with the conserved quantity varies in the two coexisting macroscopic phases and the associated chemical potential is the same in both phases. In the particular case of an isospin-symmetric system (shown in Fig.~1), the chemical potential associated with the conserved quantity charge (or isospin), $\mu_Q$, is necessarily zero. This results in an azeotropic behavior \cite{Muller:1995ji} and a necessarily congruent PT. Note that, for neutron-stars, non-congruent PT's are referred to as Gibbs constructions \cite{Glendenning:1992vb} and congruent PT's (forced by a possibly large surface tension between phases) are called Maxwell constructions. In this case, the conserved quantities involved are baryon number and zero electric charge (globally conserved if one allows for global charge neutrality). The most relevant non-congruent feature that can take place in neutron stars is the appearance of a coexistence of charged microfragments of hadrons and quarks that extends over a large portion of the star's radius (see for example Ref.~\cite{Dexheimer:2009hi}). Note that in Fig.~1 the neutron-star matter PT is congruent instead because of the assumption of local charge neutrality.

Figure~2 shows phase diagrams for isospin-symmetric matter using the CMF model. The left panel shows temperature versus baryon chemical potential and the right panel shows temperature versus pressure. In this case, there is no extended phase coexistence region and the deconfinement and confinement curves coincide. The right panel of Fig.~2 illustrates that the deconfinement coexistence line has a negative slope for $dT/dP$. This behavior is opposite from the nuclear liquid-gas PT (which is of first order in the case that Coulomb interactions are neglected). By invoking the Clausius-Clapeyron equation, we find that the negative slope stems from the fact that the entropy per baryon is larger in the quark phase than in the hadronic phase. This behavior is still different from the water solid-liquid type of PT, in which case the slope is negative but due to the fact that the density of solid water is lower than the one of liquid water in the relevant temperature range. For more details on the so called ``entropic" $dT/dP<0$ PT's, see Ref.~\cite{Iosilevskiy:2014qha}.

\begin{figure}[t!]
\includegraphics[width=7.4cm]{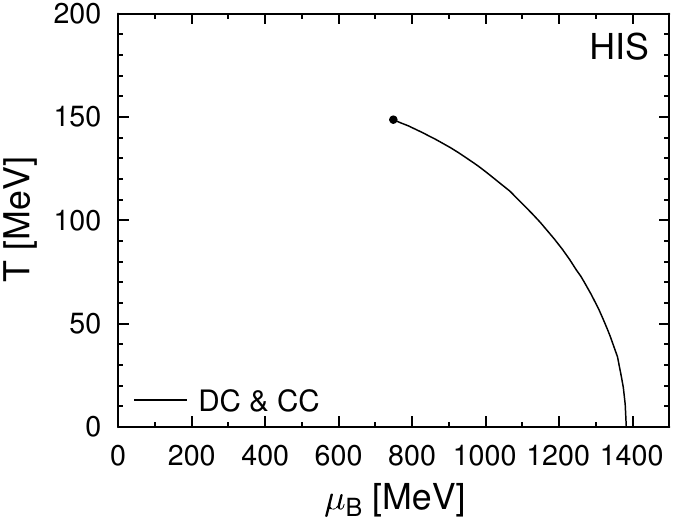}
\includegraphics[trim={0cm .25cm 0cm 0cm},width=7.5cm]{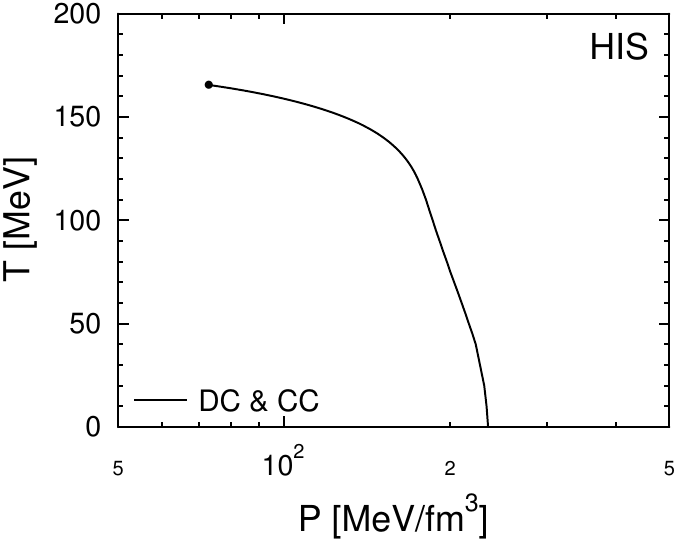}
\caption{Phase diagrams for isospin-symmetric (HIS) matter showing temperature vs. baryon chemical potential (left panel) and temperature vs. pressure (right panel). DC stands for  deconfinement curve and CC for confinement curve. In these cases they coincide.}
\end{figure}

A more interesting case pertains to lower energy heavy-ion collisions. If there is a significant net baryon density left behind when the nuclei collide, the matter created will have a charge fraction of about  $Y_Q=0.4$ in Au-Au or Pb-Pb collisions. The two conserved charges to be considered in this system are baryon number and electric charge fraction. Fig.~3 illustrates the non-congruent features that take place in the case of charge fraction $Y_Q=0.3$. The phase coexistence occupies an extended region of the diagram going from the confinement curve until the deconfinement curve. Inside, the charged chemical potential (which is the same in each phase) changes continuously, as has already been discussed in Refs.~\cite{Muller:1997tm,Shao:2011fk,Sissakian:2006dn}. Nevertheless, the area within this region becomes vanishingly small for large temperatures, as thermal effects dominate over the non-congruent features. Such an effect is not observed for the nuclear liquid-gas PT region that extends to much lower temperatures \cite{Barranco:1980zz,Muller:1995ji}. In Fig.~3, there is also a curve calculated for a forced congruent PT, in which the charge fraction is forced to be the same in both phases, which is only shown for comparison. Finally, note that the x-axis in the left panel of Fig.~3 is $\tilde{\mu}_B$, which is the chemical potential that is the same in both phases (${\mu}_B$ is not). It is defined as $\tilde{\mu}_B={\mu}_B+Y_Q\mu_Q$ and calculated from the thermodynamical potential as $d\Omega/dB$, where B is the baryon number.

In conclusion, we have presented an effective model (CMF) that has been calibrated and is suitable to describe the entire QCD phase diagram, including the description of critical points. It is in agreement with zero temperature nuclear physics, astrophysics, heavy-ion collisions, and lattice QCD. The model is consistent with perturbative QCD at zero temperature and it is being tested at the moment for finite temperature. In this work, we made use of the CMF model to study the thermodynamics of the QCD phase diagram and made comparisons between the deconfinement PT and the nuclear liquid-gas PT. We point out that only a unified equation of state description of phases (as it is usually used to study the nuclear liquid-gas PT) can provide critical points and crossovers. In addition, the assumed full miscibility of hadrons and quarks in the CMF model, e.g., in contrast to the underlying picture of simple quark-bag models, leads to the appearance of quarks embedded in the "hadronic sea" and hadrons embedded in the "quark sea". Nevertheless, quarks will always give the dominant contribution in the quark phase, and hadrons in the hadronic phase. The hadronic and the quark phase are characterized and distinguished from each other by their order parameters. We assume that the inter-penetration of quarks and hadrons in the two phases is physical, and it is required to obtain the cross-over transition at low baryon chemical potentials.

There is still much to be understood concerning non-congruent PT's and possible signatures that can be measured in heavy-ion collisions. In this work we assumed net strangeness to be locally set to zero, but extensive work has been performed assuming more complicated scenarios \cite{Heinz:1987sj,Greiner:1987tg, PhysRevD.47.2068,PhysRevD.50.4771}.

\begin{figure}[t!]
\includegraphics[width=7.4cm]{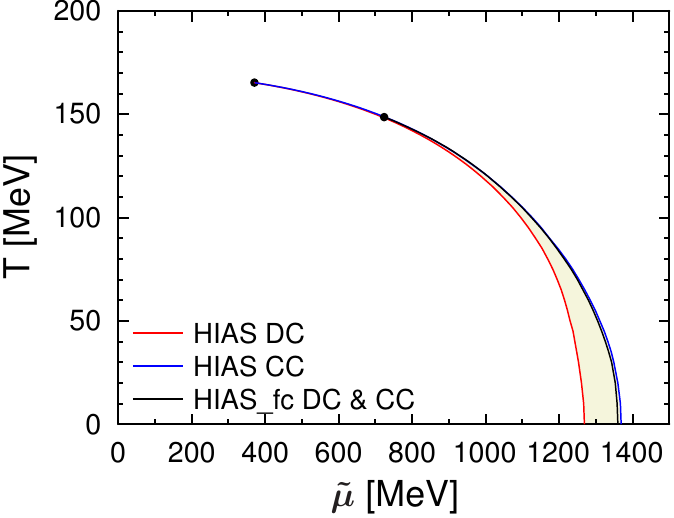}
\includegraphics[trim={0cm .25cm 0cm 0cm},width=7.5cm]{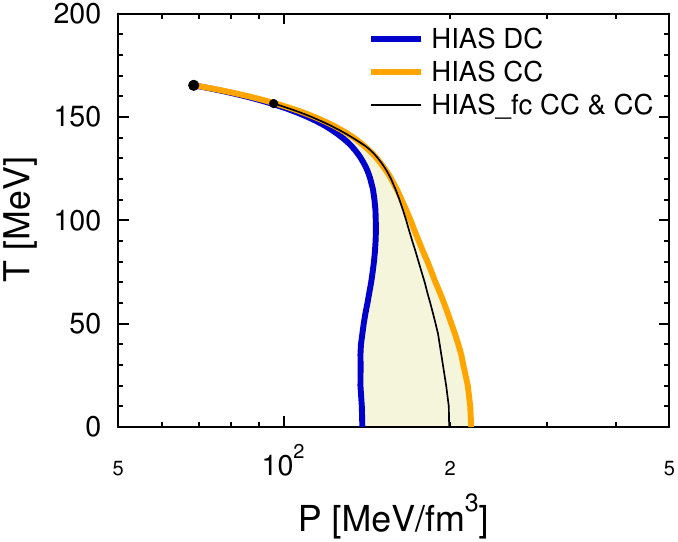}
\caption{(Color online) Phase diagrams for isospin-asymmetric (HIAS) matter with charge fraction $Y_Q=0.3$ showing temperature vs. modified baryon chemical potential (left panel) and temperature vs. pressure (right panel). DC stands for  deconfinement curve and CC for confinement curve. $f_c$ illustrates the non-physical forced congruent case.}
\end{figure}

%% References
%%
%% Following citation commands can be used in the body text:
%% Usage of \cite is as follows:
%%   \cite{key}         ==>>  [#]
%%   \cite[chap. 2]{key} ==>> [#, chap. 2]
%%

%% References with BibTeX database:

\bibliographystyle{elsarticle-num}
\bibliography{apssamp}

%% Authors are advised to use a BibTeX database file for their reference list.
%% The provided style file elsarticle-num.bst formats references in the required Procedia style

%% For references without a BibTeX database:

% \begin{thebibliography}{00}

%% \bibitem must have the following form:
%%   \bibitem{key}...
%%

% \bibitem{}

% \end{thebibliography}

\end{document}